\documentclass[prc,aps,floatfix,nofootinbib,twocolumn,letterpaper,reprint]{revtex4}
\bibpunct{[}{]}{;}{n}{}{,} 
\usepackage{graphicx}
\usepackage{color}
\usepackage{amsmath}
\usepackage{amsfonts}
\usepackage{amssymb}
\begin{document}






\title{ \vspace{1cm} Application of an {\it ab initio} $S$ matrix to data analysis of  transfer to the continuum reactions populating $^{11}$Be}
\author{ A. Bonaccorso$^{1}$, F. Cappuzzello$^{2,3}$,  D. Carbone$^{3}$, M. Cavallaro$^{3}$, G. Hupin$^{4}$,  P. Navr\'atil$^{5}$ and S. Quaglioni$^{6}$.}


\affiliation{ 
            $^1$Istituto Nazionale di Fisica Nucleare, Sezione di Pisa, Largo B. Pontecorvo 3, 56127 Pisa, Italy. \\  $^2$Dipartimento di Fisica ed Astronomia "Ettore Majorana", Universit\`a degli Studi di Catania\\  $^3$  the Istituto Nazionale di Fisica Nucleare - Laboratori Nazionali del Sud., Via S. Sofia, Catania, Italy.\\
            $^4$ Institut de Physique Nucl\'eaire, IN2P3 , CNRS, Universit\'e Paris-Sud, Universit\'e Paris-Saclay, 91406 Orsay Cedex, France\\
            $^5$ TRIUMF, 4004 Wesbrook Mall, Vancouver, British Columbia V6T 2A3, Canada.\\
            $^6$ Lawrence Livermore National Laboratory, P.O. Box 808, L-414, Livermore, California 94551, USA. }

\begin{abstract}
 Recently, the bound and continuum spectrum of $^{11}$Be has been  calculated within the {\it ab initio} no-core shell model with continuum (NCSMC) method with the parity
inversion in the ground state successfully reproduced. The continuum spectrum obtained is in agreement with known experimental levels. The $S$ matrix contained in the NCSMC continuum wave functions of  the n+$^{10}$Be system is used in  this work for the first time in a Transfer-to-the-Continuum  (TC) reaction calculation. The TC approach is applied to study the excitation energy spectrum  of $^{11}$Be measured in the  $^{9}$Be($^{18}$O,$^{16}$O)$^{11}$Be reaction at 84 MeV. Previously known levels are confirmed and theoretical and experimental evidence for a 9/2$^+$ state at E$_{x}$=5.8 MeV is given, whose configuration is thought to be $^{10}$Be(2$^+$)+n(d$_{5/2}).$
\end{abstract}

\maketitle

\section{Introduction}
This work presents the first application of an {\it ab initio} structure model for continuum states  \cite{AC} to the calculation of a transfer reaction to resonant states  \cite{bb1,bb2,bb3,bbv,noi1}. 
Thanks to the $S$ matrix obtained within the no-core shell model with continuum (NCSMC) method \cite{AC}, the traditional reduction of the complicated many-body problem to a one-body problem via a phenomenological optical potential to treat the continuum neutron-target final state is not  necessary, as previously done for $^{10}$Be \cite{noi1,bobme}. 
We will show how various states of an exotic nucleus such as $^{11}$Be are built up by the addition of one or two neutrons to a $^9$Be target.

The case of a  $^{9}$Be target is particularly interesting because one can access the $^{10}$Be and $^{11}$Be residual nuclei, which have been the subject of a large number of studies  by the nuclear physics community in the past years. 
In particular $^{11}$Be is a weakly bound nucleus with a very rich and somehow surprising phenomenology \cite{fort1,el}. Some of the  interesting characteristics  are, for example,  the role of correlations between core excited states and neutron orbitals \cite {AC,NVM,HL,paco},  the neutron-neutron pairing, and even  molecular clustering  \cite{vO}. They are all  relevant for the discussion in the present work. 
Thus by comparing one- and two-neutron transfer results, as we have consistently done for some time now,  one can understand how correlations arise in an exotic nucleus.

In our investigation we use one- and two- neutron transfer reactions induced by an $^{18}$O  projectile. In this way states in neutron-rich nuclei are populated starting from stable targets \cite{noi1}, \cite{15c,Cavallaro,Cavallaro1,Cavallaro2,Nature,Erma,Paes,Agodi}.
The observed selectivity of $^{18}$O-induced transfer reactions allows for a consistent exploration of both single particle features via the ($^{18}$O, $^{17}$O) reaction  and two-neutron correlations via the ($^{18}$O, $^{16}$O) reaction.  In particular we argue that  the ($^{18}$O, $^{16}$O) reaction  proceeds mainly via two mechanisms: (i)  a two-step single neutron transfer, where the two neutrons are independently transferred to accessible orbitals in the target field,  and  (ii)   a one-step transfer of a correlated pair of neutrons,  populating mainly two-neutron configurations in the states of the residual nucleus. In the following we  use a theoretical model corresponding to mechanism (i). 

 {Here we would like to stress the fact that in  the part of the $^{11}$Be spectrum that we aim to describe  in this work, one neutron is transferred    to a bound state while the other is transferred to the continuum thus  the independence of the two  mechanisms is better justified. At the moment there is no reaction model using  a two-neutron correlated  final state  with one neutron bound and the other unbound. 

 The   state of the art on this subject is different  for ($ t, p$) reactions populating bound states in the target. It can be summarised with the findings in Refs.\cite{GP}  and \cite{IT}, which agree with all previous literature. In the first paper  it is stated that {\it "É the simultaneous  and nonÐorthogonal contributions [of the two neutrons to transfer]  are in antiÐphase, so that the contributions corresponding to the coherent superposition of these two amplitudes  tend to cancel. The calculated total cross section thus essentially coincides with the successive  process."}
 On the other hand,  Ref. \cite{IT} shows that the two-step, successive, transfer also 
occurs in a correlated fashion. If we were to use the model in  \cite {IT},  for example, the different possible quantum 
mechanical paths associated with the population of different 
intermediate states of $^{10}$Be+$^{17}$O  in our case,  would be  summed coherently.  Also, the different terms (successive transfer, 
simultaneous transfer, and non-orthogonality contribution) would represent
different amplitude contributions to the second-order process, to be 
added coherently to get the second order cross section. Their relative 
importance in a coupled channel model such as that discussed in \cite{IT} depends on the arbitrary choice of the reaction 
representation (prior-prior, post-post, prior-post or post-prior). However, in our case because we treat transfer between heavy ions we can use  a semiclassical model in which the post and prior representations coincide. Furthermore the Transfer to the Continuum (TC) provides analytical probabilities for energy spectra in the continuum in heavy-ion reactions. Results for ($p, t$) or ($t, p$) to bound states can give only partial  guidance to this work,  due to the presence of a heavy core in the projectile and target  and of a neutron  final  continuum state in our case.}

Also for the  ($^{18}$O, $^{16}$O) two-neutron transfer reactions the 2$^+$ excited state  at 1.98 MeV in $^{18}$O plays a key role  in the coupling scheme that would require a  coupled channel theoretical treatment which at the moment is still missing \cite{Cavallaro}.  On the other hand the successful results of a semiclassical, two-step  treatment of the $^{208}$Pb($^{16}$O,    $^{18}$O)$^{206}$Pb reaction \cite{Fra}, have been used for  a long time as a justification  \cite{vOV} for neglecting interference effects and simultaneous pair transfer. In our case this is particularly true, as we do not  study the angular distribution but energy spectra.

Using the one-neutron transfer reaction we have been able in the past to obtain both bound and continuum states of $^{10}$Be \cite{noi1}. A very accurate description of the latter was achieved thanks to a previous work in which two different n+$^{9}$Be optical potentials, fitted to scattering data over a very large energy range \cite{bobme}, were compared. 
These potentials allowed some of us to calculate an $S$ matrix which is necessary for the correct definition of a continuum nucleon-nucleus state. This in turn is one of the ingredients of  the Transfer to the Continuum  model \cite{bb1,bb2,bb3} which allows us to reproduce the excitation energy spectrum of the target \cite{bbv,noi1}.
The $^{9}$Be($^{18}$O,$^{17}$O)$^{10}$Be experiment and relative theoretical description  \cite{noi1} constituted the first step in the study of    the $^{9}$Be($^{18}$O, $^{16}$O)$^{11}$Be reaction that we discuss in this paper.
Two-neutron transfer experiments on $^{9}$Be  have been performed previously at higher incident energies and with different projectiles such as tritium \cite{Liu}-\cite{tp5}, $^6$He \cite{Mayer}, $^{13}$C \cite{o1,GB2} and $^{16}$O \cite{mf}-\cite{o3}.  From the experimental point of view the novelty in our case resides in  the fact that for a heavy projectile the core spectator situation is realized, and thanks to the large relative angular momentum of the projectile and target, coupled with the initial state angular momentum, high spin states in the final nucleus can be reached \cite{Brink}.  From the theoretical point of view we  achieve here a unification of structure and reaction formalisms via the use of an {\it ab initio} $S$ matrix.

\section{The experiment}
\begin{figure}[h!]
\begin {center}
\includegraphics[scale=.35]{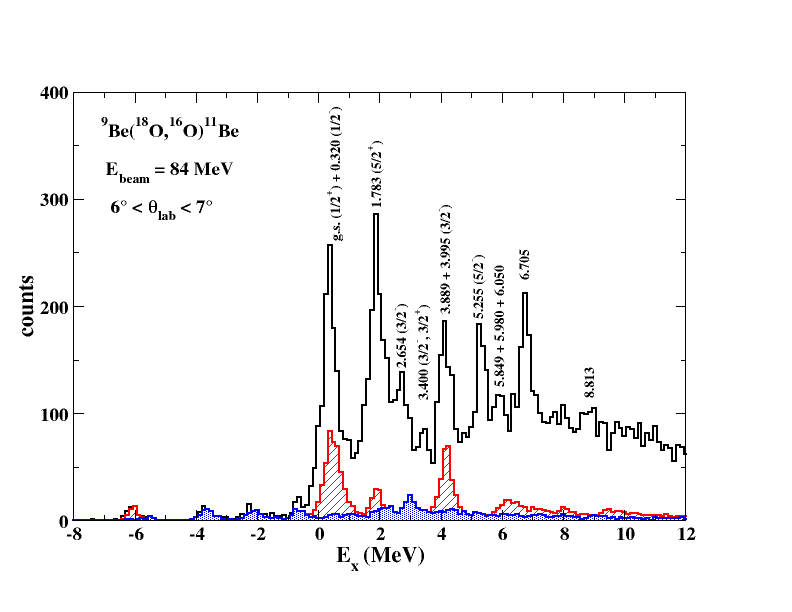}\end{center}
\caption {(Color online) Inclusive excitation energy spectrum of the $^{9}$Be($^{18}$O, $^{16}$O)$^{11}$Be reaction at 84 MeV incident energy and 6$^\circ$ $<$ $\theta_{lab}$ $<$ 7$^\circ$. The background coming from C and O contaminations in the target is shown as red and blue hatched areas, respectively.}
\label{fig1}\end{figure}

The $^{18}$O$^{6+}$ beam at 84  MeV incident energy was  accelerated by the Tandem Van de Graaff facility of the Istituto Nazionale di Fisica Nucleare - Laboratori Nazionali del Sud. A self-supporting  200 $\pm$ 10 $\mu$g/cm$^2$ thick $^9$Be  target was used. Supplementary runs with a 50 $\pm$ 3 $\mu$g/cm$^2$ self-supporting $^{12}$C target and a 260 $\pm$ 10 $\mu$g/cm$^2$ WO$_3$ target were recorded in order to estimate the background in the energy spectra from $^{12}$C and $^{16}$O impurities in the $^9$Be target. The MAGNEX magnetic spectrometer \cite{magnex_rev} was used to momentum analyze the $^{16}$O ejectiles, detected by its focal plane detector \cite{FPD}. The angular range between 3$^\circ$ and 10$^\circ$ in the laboratory reference frame was explored. Details about the particle identification and the trajectory reconstruction techniques used for the reduction of the MAGNEX data can be found in Refs. \cite{pid,epjp,eff,momento}.  An overall energy and angular resolution of about 500 keV (FWHM) and 0.3$^\circ$ were obtained. The absolute cross section was also estimated according to Ref. \cite{epjp}, with a total error of about 10\% induced by the uncertainties in the target thickness and beam current integration. 
An example  of an energy spectrum for the ($^{18}$O, $^{16}$O) reaction with a $^{9}$Be target, in a limited angular range,   is shown in Fig. \ref{fig1}, where the background contributions are also shown.
The angle integrated absolute cross-section  spectrum obtained after  background subtraction is shown in Fig. \ref{fig2}.

\section{Model independent analysis of the energy spectrum}
A number of peaks signals the population of bound and resonant states of $^{11}$Be in this transfer reaction. 

In our data the 1/2$^+$ ground state and the 1/2$^-$ first excited state at 0.320 MeV are not resolved. Above  the one-neutron separation energy $S_n$ = 0.502  MeV, a strong excitation of the 5/2$^+$ state at 1.783 MeV is observed, while the  states at 2.654 and 3.400 MeV are less populated. 
Interestingly these  are the only states observed in one-neutron transfer reactions in the past \cite{TKS}, \cite{DA}, \cite{BZ}, showing a  pattern  similar  to that observed here. This could indicate that, for these states, the $^{10}$Be$_{gs}$ is preferentially populated as an intermediate system in the  first step of the reaction and a second neutron is then transferred to accessible single-particle orbitals. 

Beyond the $^{10}$Be core excitation threshold we observe a peak at 3.9 MeV, where we cannot separate the transitions to the known states at 3.889 MeV (5/2$^-$) and 3.955 MeV (3/2$^-$). Only in Ref. \cite{Liu} were the two states separated in a two-neutron transfer reaction, showing that they are both strongly excited by such a probe. Intense transitions are observed for the 5/2$^-$ state at 5.255 MeV and the state at 6.705 MeV, resembling the situation observed in all the reported two-neutron transfer reaction data \cite{o1}, \cite{GB2}, \cite{mf}. These states are not populated in single-neutron transfer or in the $^{11}$B($^7$Li, $^7$Be)$^{11}$Be charge exchange reaction \cite{FC2001}, \cite{FC2004}, suggesting a selective response of $^{11}$Be to two-neutron transfer operators, likely populating $^9$Be$\otimes$(sd)$^2$ configurations. 
Studies of  $\beta$ decay  from the 3/2$^-$ $^{11}$Li parent nucleus show transitions to the 3/2$^-$ state at 3.955 MeV and 5/2$^-$ at 5.255 MeV of $^{11}$Be. This does not contradict the above argument about the $^9$Be$\otimes$(sd)$^2$ configurations for these states, since $\beta$ decay can also occur within the $^9$Li core of the $^{11}$Li parent nucleus. In addition, $\beta$ decay does not populate the state at 6.705 MeV, which would indicate either a different parity or a  spin higher than that achievable by the allowed Fermi and Gamow-Teller operators. In Refs.  \cite{o1} - \cite{mf} this state is described as belonging to a rotational band built on the 3/2$^-$ state at 3.955 MeV, together with the 5.255 MeV state, with a 7/2$^-$ spin assignment. 

Also debated are the states at 5.8  and 8.813 MeV, which are both present in all reported two-neutron transfer experiments, even though the centroid energy and width are slightly different in the various studies. 
The former is not observed in $\beta$ decay or in single neutron transfer, but it is likely the same as observed in  the$^{11}$B($^7$Li,  $^7$Be)$^{11}$Be charge exchange reaction at 6.05 MeV (FWHM 300keV) \cite{FC2001}. Due to the large momentum transfer, such a reaction can easily excite states with a high spin, not populated in $\beta$ decay. In Ref. \cite{mf} the neutron decay of this state to the first excited 2$^+$ state of $^{10}$Be suggests a core-excited configuration. Recently a 9/2$^+$ spin-parity for a state with a [$^{10}$Be$(2_1^+)$$\otimes$(1d)$^{5/2}]^{9/2+}$ stretched configuration in the  $^{10}$Be$(2_1^+)$ excited state was predicted  in the {\it ab initio} NCSMC approach~\cite{AC}, which is  in perfect agreement with the present finding and our interpretation of previous literature. 
For the state at 8.813 MeV, $\beta$-decay studies \cite{Hira, Fin} assign a 3/2$^-$ spin parity,  conflicting with the assumption of a high-spin member of the rotational band built on the 3/2$^-$ state at 3.955 MeV proposed in Refs. \cite{mf}- \cite{GB2}. Our data, despite confirming the population of this state in two-neutron transfer reactions, do not add much to the previous debate.

 \begin{table} [h]
\caption{Structure parameters of the valence neutron (see text.) The  $^{10}$Be spectroscopic factors in the present work were obtained with the same chiral N$^2$LO$_{\rm SAT}$ NN+3N interaction~\cite{n2losat} as used in the $^{11}$Be calculations. The $C^2S $ and $C_i$ values for  $^{17}$O have  been used in the calculations of Eq.(\ref{eq3}) while the others are given here  for completeness.}
\centering  
\begin{ruledtabular}
\begin{tabular} {cccccc}  
\noalign{\smallskip}
State&$S_{n}$ & $ C_i$ &$j$ &$l$&$C^2S $ \\
&(MeV)&(fm$^{-{1\over2}}$)&&&\\
$^{18}$O$_{g.s.}$&8.04& 1.73&$5/2$&2& 1.7\cite{Cavallaro} \\
$^{17}$O$_{g.s.}$&4.145 &0.69& 5/2&2&0.945\cite{Cavallaro}\\
$^{10}$Be$_{g.s.}$&6.812 &2.3 &3/2 &1&2.13\cite{VMC}, 2.6\cite{sf}, 2.73\footnote{From the present work.} \\ 
$^{10}$Be$(2_1^+)$&3.44& &  1/2 &1&0.0226\cite{VMC}, 0.274\cite{sf}\footnote{Reference \cite{sf} does not distinguish the two $j$-components.}, 0.036$^{\rm a}$  \\ 
$^{10}$Be$(2_1^+)$&& &  3/2 &1&0.268\cite{VMC}, 0.274\cite{sf}$^{\rm b}$, 0.24$^{\rm a}$   \\ 
$^{10}$Be$(2_2^+)$&0.854&&1/2&1&0.616\cite{VMC}, 0.421\cite{sf}$^{\rm b}$, 0.406$^{\rm a}$  \\ 
$^{10}$Be$(2_2^+)$&&&3/2&1&0.13\cite{VMC}, 0.421\cite{sf}$^{\rm b}$, 0.076$^{\rm a}$  \\ 
\noalign{\smallskip}
\label{t1}
\end{tabular}
\end{ruledtabular}
\end{table}

\section{Theoretical approach}
From the theoretical point of view, the description of the n + $^{10}$Be part of the spectrum above $S_n$ is very challenging because no experimental data exist on the free n + $^{10}$Be scattering. 

However, the recent {\it ab initio} calculation of  the $^{11}$Be spectrum \cite{AC} over a wide energy range, allows for a description of both bound and continuum states, the continuum part providing an $S$ matrix. Thus we can apply the TC method overcoming the knowledge of the optical potential thanks to the NCSMC $S$ matrix. One particularly interesting aspect is that the n + $^{10}$Be component of the spectrum has contributions from bound states of  $^{10}$Be. 

\subsection{Reaction model:}
We imagine the reaction going through two independent steps as anticipated in Sec. I and indicated as mechanism (i). In the first step the reaction $^{9}$Be($^{18}$O, $^{17}$O)$^{10}$Be takes place and neutron n1 is transferred from $^{18}$O to populate the bound states of $^{10}$Be.

We then imagine that the second neutron, n2, is transferred from $^{17}$O to $^{10}$Be, and since we just want to describe  the  continuum part of the spectrum (above $E_{x}=S_n$)  we use  the following   formula for the two-neutron transfer description:

\begin{equation}{d \sigma _{2n} \over  {d\varepsilon_{f2}}}=C^2S~
  P_{n1}^{phen}(R_s)  \int_{b_{min}}^ {b_{max}}d{\bf b_{c}    }   |S_{ct}(b_{c}    )|^2  {dP_{n2}(b_c    )\over d{\varepsilon_{f2}}} ,\label{eq3}\end{equation}
  $ P_{n1}^{phen}$, given by Eq.(\ref{eq2}), is the bound-state  transfer probability for neutron n1 from $^{18}$O to  $^{9}$Be, at the strong absorption radius between core and target, defined as R$_s$=1.4(A$_c^{1/3}$+A$_t^{1/3}$). Then in the next step, n2 is the second neutron which is transferred to a continuum state from $^{17}$O and
  $C^2S$ is its initial wave function  spectroscopic factor. $\varepsilon_{f2}$ is the continuum final-state energy of n2 and $|S_{ct}(b_{c})|^2 $ has been calculated according to  \cite{bcc}. It is the elastic scattering probability between $^{16}$O and $^{10}$Be, the second-step core  and target nuclei, respectively.   Following Ref. \cite{noi1}, the correspondence between the measured scattering
angle 3$^\circ$ $<$ $\theta_{lab}$ $<$ 10$^\circ$ and the impact parameter has been  obtained via a classical trajectory calculation according to Ref. \cite{BW}, providing
$7 < b_c < 8$  fm to be used in Eq.(\ref{eq3}).  $P_{n1}^{phen}$ is extracted as described below, from the integrated experimental one-neutron transfer cross section in the $^{9}$Be($^{18}$O, $^{17}$O)$^{10}$Be reaction whose  data were presented in \cite{noi1}.  Above the threshold for the first excited state of $^{10}$Be,  $ P_{n1}^{phen}$ is obtained consistently using   the parameters  appropriate to take into account the excitation energy in $^{10}$Be listed in Table \ref{t1}.

 To take advantage of the fact that the experimental cross section for one-neutron transfer is already known  and to minimize the dependence from the parameter's incertitude, we use  the one-neutron transfer cross section between bound states formula from  \cite{bb0},
 \begin{equation}{ \sigma _{1n} }=
 \pi{(R_s-a_c)\over \eta} P_{n1}(R_s)  \label{eq1}\end{equation}
 which is obtained when the core-target $S$ matrix is calculated in the sharp cut off approximation:  $$|S_{ct}({ b}_{c}  )|^2=1~~~~~~~~ {\rm if }~~~~~~~~~~ b_c>R_s$$
   $$|S_{ct}({ b}_{c}  )|^2=0~~~~~~~~ {\rm if }~~~~~~~~~~ b_c<R_s.$$ $\eta$ is a kinematical parameter  depending on the initial and final neutron separation energies and the energy of relative motion \cite{bb0},
   
    $$\eta=\sqrt{\gamma_i^2+k_1^2}$$ where $ \gamma_i^2=2mS_{i(1,2)}/\hbar^2$ and $S_{i(1,2)}$ is the bound  state initial separation energy of the first and second neutron, respectively. $ k_1^2=(Q+1/2mv^2)^2/(\hbar v)^2$  where $Q=\varepsilon_{i(1,2)}-\varepsilon_{f(1,2)}$ is  the $Q$ value.  The $\varepsilon_{i(1,2)}$ are the negative initial binding energies of neutrons n1 and n2 in their bound states, while $\varepsilon_{f(1,2)}$ is the negative final energy of the first-step neutron in $^{10}$Be and is the positive continuum energy of neutron n2 in $^{11}$Be. The $\eta$ value ranges from 0.68  to 0.88fm$^{-1 }$ for the three bound  $^{10}$Be states considered here.

    R$_s$ is again the strong absorption radius and $a_c$ is the Coulomb length parameter. Then we derive from Eq.(\ref{eq1}) the  following phenomenological transfer probability at the strong absorption radius to be used to describe the two-neutron transfer in Eq.(\ref{eq3}):
  \begin{equation}P_{n1}^{phen}(R_s) = {{ \sigma _{1n}^ {exp}\over {
 \pi{(R_s-a_c)\over \eta}}} } \label{eq2}\end{equation}
where  $\sigma _{1n}^ {exp}$=
2.1 mb ($^{10}Be_{g.s.}$),
2.2 mb (${^{10}Be_{2_1^+}}$) and
3.2 mb (${^{10}Be_{2_2^+}}$) are  the experimental values from Ref. \cite{noi1}. By adopting this formula we assume that all structure information, such as spectroscopic factors,  of both projectile ($^{18}$O) and target ($^{9}$Be) are included in the measured $\sigma _{1n}^ {exp}$. 

 Our approach is equivalent to the treatment of a two-step process via second-order perturbation theory \cite{pot1,BW}, which is justified by the small first-step average probability ($P_{n1}^{phen}$=0.01; cf. Fig. 2). Furthermore, the case of $^{10}$Be, as an intermediate system in the transition from $^{9}$Be to $^{11}$Be, is quite peculiar and makes our assumption of incoherent summation still approximately valid. The key point is that the $^{10}$Be level density is quite low in the energy region of interest, and in fact we only need to consider three states, i.e. $^{10}$Be$_{g.s.}$ and the excited states at 3.4 MeV ($^{10}$Be$_{{2_1}^+}$)  and 6 MeV ($^{10}$Be$_{{2_2}^+}$). As a consequence, the $^{10}$Be$_{{2_1}^+}$  has an important role in $^{11}$Be spectra  from the inelastic thresholds of ($^{10}$Be$_{{2_1}^+}$)+n, located at about 3.9 MeV, while the $^{10}$Be$_{{2_2}^+}$ from the inelastic thresholds of ($^{10}$Be$_{{2_2}^+}$) + n, located at about 6.5 MeV. Thus for $^{11}$Be states up to 3.9 MeV the role of $^{10}$Be first and second excited states is expected to be weak. For states of $^{11}$Be between 3.9 and 6.5 MeV one can have the contribution from $^{10}$Be$_{g.s.}$ and the first $^{10}$Be$_{2^+}$. However, the states of $^{11}$Be in that energy range  with a pronounced $^{10}$Be$_{g.s.}$ + n configuration are quite broad since  they are  far from the ($^{10}$Be$_{g.s.}$)+n threshold, located at S$_n$ = 0.5 MeV. As an example, a simulated pure single-particle d$_{3/2 }$ state at 4 MeV excitation energy would have a width as large as 3 MeV or more. Therefore the role of such configurations, so  spread out in energy, is minor in the narrow region covered by the analyzed sharp resonances. In fact, the 9/2$^+$ state we claim at 5.9 MeV is built with the  $^{10}$Be first excited state  and not with the $^{10}$Be$_{g.s.}$, as the ab initio calculation of \cite{AC}  correctly indicates. The same argument holds for $^{11}$Be states above 6.5 MeV where configurations with $^{10}$Be$_{g.s.}$ and the first  $^{10}$Be$_{2^+}$   likely do  not   contribute  significantly.

The second neutron, n2, transfer probability to the continuum states of $^{11}$Be  is given by
\begin{widetext}
 \begin{equation}   {dP_{n2}(b_c    )\over d{\varepsilon_{f2} }} ={ |C_i|^2\over
2 k_{f2}} \left [{\hbar\over mv}\right ]{e^{-2\eta b_c    }\over 2\eta b_c } \Sigma_{j_f,\nu}(|1-\bar S_{j_f,\nu} |^2+1-|\bar S_{j_f,\nu} |^2)
(2j_f+1)(1+R_{if})M_{l_fl_i}. \label{eq4}\end{equation}
\end{widetext}
Details of the TC method have been given in several previous publications; see, for example Ref. \cite{bb2}, where the definitions of the parameters appearing in Eq. (\ref{eq4}) can also be found. 

 In Eq. (\ref{eq4}) for each continuum energy the sum is over all possible n+$^{10}$Be total angular momenta. Above the thresholds for the first and second $2^+ $ excited states in $^{10}$Be, for each angular momentum there is also an incoherent sum over all channels $\nu$ contributing to it. This is the same as when calculating total-reaction cross sections and the  observable is the final nucleus excitation energy spectrum. The situation is different (c.f. Eq.(6) in Ref. \cite{pot1}) when two- neutron transfer is discussed with the aim of calculating  core angular distributions. In the latter case for each angle different channels can interfere and contribute coherently. 
 
  We have usually calculated $\bar S_{j_f}$ from an optical potential \cite{bb2}. The same procedure cannot be applied to the system n+$^{10}$Be because there are no data available. However the $S$ matrix calculated by some of us in Ref. \cite{AC} is perfectly suited to be used in Eq. (\ref{eq4}) as it is given in terms of the n-$^{10}$Be  continuum energy and  angular momentum $j_f$ appearing in Eq. (\ref{eq4}) and it contains, at each energy, all possible inelastic channel contributions. Note that Eq. (\ref{eq4}) contains two terms  proportional to $|1-\bar S_{j_f} |^2$ and $1-|\bar S_{j_f} |^2$, giving the elastic and inelastic neutron breakup from the initial state in the projectile, respectively.

\subsection{Ab initio $S$ matrix}

Some of us have investigated the structure of $^{11}$Be by studying the $^{10}$Be+n system within the NCSMC approach~\cite{AC}. This approach~\cite{Na16} uses a basis expansion with two key components: one describing all nucleons close together, forming the $^{11}$Be nucleus, and a second  describing the separated neutron and $^{10}$Be clusters. The former part utilizes a square-integrable harmonic-oscillator basis expansion treating all 11 nucleons on the same footing. The latter part factorizes the wave function into products of $^{10}$Be and neutron components and their relative motion with proper bound-state or scattering boundary conditions. We compute the $S$ matrix from the 11-body NCSMC calculations by employing the calculable $R$ matrix method \cite{55,56} and matching the many-body $^{11}$Be internal wave function with the asymptotic binary n-$^{10}$Be channels at  about 18 fm, well beyond the range of the nuclear interaction. The chiral N$^2$LO$_{\rm SAT}$ two-nucleon (NN) and three-nucleon (3N) interaction~\cite{n2losat} served as input. For the $^{10}$Be cluster, in addition to the ground state, we also included the first and the second excited $2^+$ states. The outcomes of the NCSMC calculations are the energies and wave functions of the bound states, here $1/2^+$ and $1/2^-$ in the correct order compared to experiment, as well as of the continuum states. The latter include the $S$ matrix that we in turn apply in the present investigation of the $^{9}$Be($^{18}$O, $^{16}$O)$^{11}$Be two-neutron transfer reaction. As shown in Figs. 2 and 3 and Table I in Ref.~\cite{AC}, the N$^2$LO$_{\rm SAT}$ interaction provides a quite reasonable description of the low-lying bound states and resonances of $^{11}$Be. However, to achieve the spectroscopic accuracy needed in the present study, we turn to the NCSMC-pheno approach which allows us to reproduce the experimental thresholds and bound- and resonant-state energies as presented in the right part of Table I in Ref.~\cite{AC}. The NCSMC calculations  with the N$^2$LO$_{\rm SAT}$ interaction predict a low-lying $9/2^+$ resonance at 5.42 MeV and, with the NCSMC-pheno approach, at 5.59 MeV after adjusting the $^{10}$Be thresholds to experiment. In this work we examine an experimental candidate for this resonance at 5.3 MeV. Note that these energies are given with respect to the n+$^{10}$Be threshold. Consequently, we have performed a new NCSMC-pheno calculation to fit the calculated $9/2^+$ resonance position to that energy. The resulting $S$ matrix is then used in the present study.

\begin{figure}[h!]
\begin {center}
\includegraphics[scale=.35]{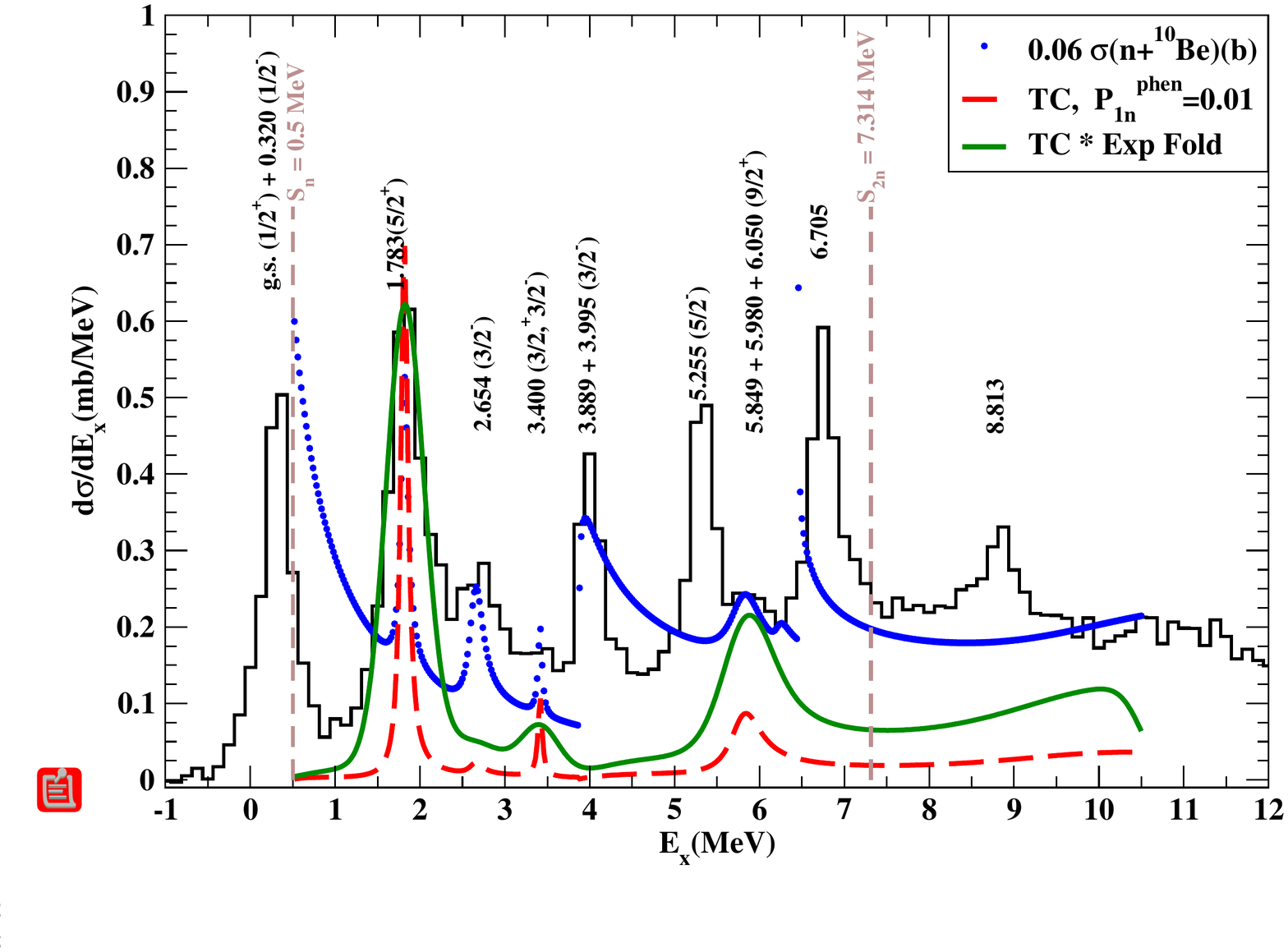}\end{center}
\caption { (Color online) $^{11}$Be inclusive excitation-energy spectrum from the $^{9}$Be($^{18}$O, $^{16}$O)$^{11}$Be reaction at 84  MeV incident energy, integrated in the total measured angular range 3$^\circ$ $<$ $\theta_{lab}$ $<$ 10$^\circ$. Black histogram: experimental data after background subtraction.  Dashed  red curve: total 2n transfer calculations resulting from the use of the $S$ matrix from Ref. \cite{AC}. Solid green curve: the red curve calculation folded with the experimental resolution and renormalized by  a factor  of 3.2. Dotted blue curve: theoretical free n+$^{10}$Be cross section, in barn, see text for details.}
\label{fig2}\end{figure}

\section{Discussion of the results }
 In Fig. \ref{fig2} we show the experimental  inclusive energy spectrum of $^{11}$Be together with the theoretical calculations (dashed red curve). This calculation does not contain any fitting parameter.   On the other hand the solid green  curve is the same calculation folded with the experimental resolution renormalized by  a factor of 3.2 to fit the data.  The first three continuum states are well reproduced  as far as the position is concerned, indicating that the TC model contains the correct dynamics and the {\it ab initio} $S$ matrix accurate structure information. Note that also the order of  magnitude is reproduced reasonably well within the incertitude due to the description of the two-neutron transfer reaction mechanism. 
The calculated relative population of the three resonances at E$_{x}$ = 1.783, 2.654 and  3.400 MeV  with $J^\pi=5/2^+, 3/2^- $ and  $3/2^+$, respectively, 
compares well with the experimental results for $^{10}$Be($d, p$)$^{11}$Be \cite{TKS}. The state at 2.654 MeV is depleted in our model calculation because of the unfavorable  one-neutron transfer reaction matching, as it is in the data from \cite{TKS}. However, it is well seen in the present data as well as in all previous two-neutron transfer experiments \cite{mf}, confirming for this state a relevant 2n+$^9$Be configuration and a population via the simultaneous 2n-transfer reaction mechanism, type (ii) discussed in Sec.I.  On the other hand the calculated  peak at E$_{x}$ = 3.400 MeV, has such a small width ($\approx $0.02 MeV)  from \cite{AC} that its presence  has little physical significance and in fact it is hardly visible as a structure in the experiment. The transfer calculation probably misses the maximum value for the same reason. Note that in the figures we indicate ($3/2^+$, $3/2^-$) for this state   because from the literature it is  often quoted as $3/2^-$, while from the present $S$ matrix calculations it appears to be $3/2^+$.   As far as the absolute total cross sections are concerned we believe that  due to the experimental and theoretical incertitudes the only comparable values are those for the  1.783 MeV d$_{5/2}$ state for which we get 	$\sigma_{exp}  =  359  \pm 35 (\mu b)$ and $\sigma_{th}  =  175  (\mu b)$.  However, one might argue that due to the indistinguishability of the two neutrons, Eq. (\ref{eq3}) should contain a factor of 2 which would give an almost perfect agreement between theory and experiment. On the other hand, as we have discussed before our transfer mechanism model is over-simplified and thus we conclude that the present absolute theoretical cross sections are reliable only at the order -of-magnitude level.
 
 The cross-section of the resonance at 5.8 MeV is reasonably well reproduced by our model, thus strengthening the interpretation of a 9/2$^+$ stretched configuration, discussed above and in Ref. \cite{AC}. It is the only state that can have a definite single-particle nature in the high-energy part of the spectrum, because it has an high spin which does not mix with underlying components of low angular momenta. It can be seen only in a reaction with heavy ions where the matching conditions allow the reaching of a high resonance energy and high angular momentum. This state can be reproduced  by the TC, which contains explicitly the spin couplings between initial and final states besides the angular momentum couplings. Therefore  the present work provides the first  evidence for the existence of such a state. 
 We believe our attribution is quite firm and well justified and is unique so far in the literature. From the point of view of the structure part, we can see the 9/2$^+$ because it is built on the $2^+_1$ state of $^{10}$Be that we include in our NCSMC calculation and thus it comes about as a 9/2$^+$ from the $S$ matrix calculation. 
 
Finally, for comparison,  the free n+$^{10}$Be cross section (elastic plus inelastic) calculated with the {\it ab initio} $S$ matrix and rescaled by a factor of 0.06, is shown by the blue dots in Fig. \ref{fig2}.  For this cross section the $y$-axis scale should be read in units of barns as indicated in the caption. 
Obviously this is another observable with respect to the transfer cross section shown in the figure. 
Note that the magnitude of this cross section is consistent with  that of the n+$^{9}$Be data and calculation in Ref. \cite{bobme}. This is a proof of the accuracy of the {\it ab initio} model in providing the magnitude of the neutron-core couplings (interaction). The three resonance states coupled to the $^{10}$Be$_{g.s.}$ indeed appear  at the same positions as in the data and they scale as the simple $1/{k}^2$ law. Above the first and second 2$^+$ excited-state thresholds, respectively, the shown free cross section is  the sum of the tails of the cross sections from lower neutron energies, in accordance with the hypothesis that $^{10}$Be excited states are populated by the first neutron.

\section{Conclusions}
In conclusion we have presented new experimental data on the $^{9}$Be($^{18}$O, $^{16}$O)$^{11}$Be two-neutron transfer reaction at 84 MeV, providing the absolute cross section energy spectra.
 An {\it ab initio} $S$ matrix coupled for the first time with  the TC method is able to reproduce the position and widths of the n + $^{10}$Be components in the experimental  excitation energy spectrum of the exotic nucleus $^{11}$Be and to predict the order of magnitude of the absolute cross sections. 
We have given evidence, both theoretical and  experimental, of the presence of a 9/2$^+$ state at  E$_{x}$=5.8 MeV with a  [$^{10}$Be$(2_1^+)$$\otimes$(1d)$^{5/2}]^{9/2+}$ configuration.

It is very encouraging to see that very elaborate {\it ab initio} structure calculations for continuum states can provide the correct ingredients to be easily incorporated in a reaction model which is simple and whose accuracy is within the present state of the art of the literature. 
We have also shown that exotic nuclei can be successfully  studied at stable beam facilities providing interesting complementary information to that obtained with Radioactive Ion Beams.

\acknowledgments

A.B. is grateful to J. Dohet-Eraly for enlightening discussions on the structure of the {\it ab initio} $S$ matrix in the early stages of this work.
Two of us, D.C and M.C., received funding from the European Research Council  under the European UnionÕs Horizon 2020 research and innovation program (Grant Agreement No. 714625).
P.N.'s  work was supported in part by  NSERC Grant No. SAPIN-2016-00033. TRIUMF received federal funding via a contribution agreement with the National Research Council of Canada. Computing support for G.H., P.N. and S.Q.  came from an INCITE Award on the Titan supercomputer of the Oak Ridge Leadership Computing Facility Oak Ridge
National Laborator, from Compute Canada, and from the Lawrence Livermore National
Laboratory (LLNL) institutional Computing Grand Challenge Program.
This article was prepared in part by LLNL  (S.Q.) under Contract No. DE-AC52-07NA27344. This material is based in part upon work supported by the U.S. Department of Energy, Office of Science, Office of Nuclear Physics, under Work Proposal No. SCW0498.

\end{document}